\documentclass[epj,twocolumn]{webofc}
\usepackage[varg]{txfonts}   
\usepackage{bm}
\renewcommand{\vec}[1]{\bm{#1}}
\usepackage{microtype}
\woctitle{Powders \& Grains 2017}
\begin{document}
\title{Kinetic theory for strong uniform shear flow of granular media at high
density}
%
%

\author{\firstname{W. Till} \lastname{Kranz}\inst{1}\fnsep\thanks{\email{till.kranz@dlr.de}} \and
        \firstname{Matthias} \lastname{Sperl}\inst{1,2}}
\institute{Institut für Materialphysik im Weltraum, Deutsches Zentrum für Luft- und Raumfahrt e.V., Linder Höhe,
  51147 Köln, Germany 
\and
Institut für Theoretische Physik, Universität zu Köln, Zülpicher Straße 77,
50937 Köln, Germany
          }

\abstract{%
  We discuss the uniform shear flow of a fluidized granular bed composed of
  monodisperse Hertzian spheres. Considering high densities around the glass
  transition density of inelastic Hertzian spheres, we report kinetic theory
  expressions for the Newtonian viscosity as well as the Bagnold
  coefficient. We discuss the dependence of the transport coefficients on
  density and coefficient of restitution.  
}
\maketitle

\section{Introduction}
\label{sec:introduction}

Flows of dense granular materials are ubiquitous in nature and industry
\cite{hutter+kirchner13,ausloos+lambiotte05,iverson97,lissauer93,jaeger+nagel92}. For
efficiency \cite{kunii+levenspiel91} and disaster prevention \cite{brunsden99}
a solid understanding of the rheology of these flows is crucial. Inclined
plane flows both in industrial and geophysical settings have been successfully
modeled by inelastic Hertzian spheres
\cite{jaeger+nagel92,campbell06,silbert+ertas01,cheng+lechman06,rycroft+grest06}
using molecular-dynamics simulations. Discrete element models are constrained,
however, to relatively small system sizes due to computational demands.

For a continuum modeling which can be applied at very large scales and a
theoretical analysis, it is essential to have access to the transport
coefficients as functions of the system parameters. For low densities, the
kinetic theory of granular flows is well developed based on the inelastic
Boltzmann equation
\cite{brilliantov+poeschel10,poeschel+luding01,garzo+montanero02}. It yields
predictions for, amongst others, the diffusivity and the shear viscosity. For
high densities and especially away from the linear response regime, results
derived from first principles are rare.

The Integration Through Transients (\textsc{ITT}) formalism has been
very successful in describing shear in colloidal suspensions at
densities around the glass transition and for finite shear rates
\cite{fuchs+cates02,brader+cates08,fuchs+cates09}. A glass transition
is observed \cite{abate+durian06} and predicted for fluidized granular
beds also \cite{gmct10,gmct13}. This allowed us recently to generalize
the ITT formalism to the far from equilibrium regime of a driven
granular medium \cite{kranz+frahsa16}. In this contribution we will
present results for the viscosity and Bagnold coefficient. Hayakawa
\textit{et al.} have derived two related sets of equations
(Refs.~\cite{hayakawa+otsuki08} and \cite{suzuki+hayakawa15}) where
only the second version has yielded numerical results for the
transport coefficients so far.

The structure of the paper is as follows. In Sec.~\ref{sec:model} we
will define the model and control parameters. A brief sketch of the
theory and the required inputs are given in
Sec.~\ref{sec:analytical-theory} before we will discuss the transport
coefficients for the limiting cases of weak
(Sec.~\ref{sec:linear-rheology}) and strong
(Sec.~\ref{sec:nonlinear-rheology}) shear. We will close with a
discussion in Sec.~\ref{sec:discussion}.

\section{Model}
\label{sec:model}

We consider a monodisperse system of $N\to\infty$ inelastic Hertzian
spheres \cite{pamies+cacciuto09,landau+lifshitz86} of diameter $d$ and
mass $m=1$ in a volume $V$ such that the density $n = N/V$ remains
finite. The overlap potential is given by
\begin{equation}
  \label{eq:1}
  V(r_{ij}) = \Gamma T(1 - r_{ij}/d)^{5/2}\quad\text{for } r_{ij} < d
\end{equation}
where $\Gamma$ controls the strength of the repulsion and
$T = \langle\vec v^2\rangle/3$ is the granular temperature. The
distance between two particles is denoted by
$r_{ij} = |\vec r_i - \vec r_j|$. Here we will focus on the
practically relevant case of rather stiff particles,
$\Gamma\gg1$. Dissipation is quantified by a constant coefficient of
restitution $0 \le \epsilon < 1$ \cite{brilliantov+poeschel10}.

We consider the quiescent system to be in a fluidized state
\cite{ojha+lemieux04,abate+durian06,schroeter+goldman05} and model the
fluidization by a Gaussian random force $\vec\xi_i(t)$ with zero mean
$\langle\vec\xi_i\rangle = 0$ acting on each particle
individually. Its variance
$\langle\xi_i^{\alpha}(t)\xi_j^{\beta}(t')\rangle =
2P_D\delta_{ij}\delta^{\alpha\beta}\delta(t - t')$
is characterized by the driving power $P_D$.

We imagine that at time $t=0$ the system is in a stationary state where the
dissipation is balanced by the driving force resulting in a stationary
temperature $T_0$. At this point we switch on a uniform linear shear profile
modeled by the \textsc{SLLOD} equations of motion \cite{morriss+evans07}. We
prescribe the shear rate $\dot\gamma$ and monitor the resulting shear stress
$\sigma$. For finite shear rates the shear will contribute to the energy input
by a term $\dot\gamma\sigma$.

If we wait long enough the system will reach a new stationary state with a new
temperature $T$ and constant mean shear stress $\sigma$. This is the state we
want to consider in the following. For a given material of the particles
prescribed by the pair ($\epsilon$, $\Gamma$), the macroscopic state of the
system is fully specified by the packing fraction $\varphi = \pi nd^3/6$, the
temperature $T_0$ (or, equivalently, $T$) resulting from the energy balance,
and the shear rate $\dot\gamma$.

\section{Calculating the Transport Coefficients}
\label{sec:analytical-theory}

The transport coefficients which have traditionally been calculated from the
Boltzmann-Enskog equation using, e.g., the Chapmann-Enskog or related methods
\cite{chapman+cowling70}, are necessarily valid only to lowest order in the
shear rate $\dot\gamma$ and the packing fraction $\varphi$. To go beyond the
lowest order presents a considerable challenge.

For densities around the (dynamic) glass transition, the ITT formalism enables
to calculate corrections beyond the linear order in packing fraction and for
arbitrary shear rates. It relates expectation values in the stationary sheared
state to the (equilibrium) unsheared state and captures the transient dynamics
after the switch-on of shear in terms of time-dependent density correlation
functions \cite{fuchs+cates09}. For the latter, mode-coupling theory
(\textsc{MCT}) \cite{goetze08} affords a faithful description for the
densities under consideration.

Granular \textsc{MCT} allows for a generalization of the ITT formalism for
dissipative particles. A crucial input for both \textsc{MCT} and \textsc{ITT}
are static structure factors, $S_k$, in the quiescent state. To this end we
use numerical solutions of the Hypernetted-Chain-Equation
\cite{martynov+sarkisov83} for Hertzian spheres \cite{heinen+sperl16}.

For the low-density (bare) transport coefficients, we use the Enskog
expressions provided by Garzó and Montanero
\cite{garzo+montanero02}. Especially important are the shear viscosity
$\eta_0$ and the related sound-damping constant $D_S$.

\section{Linear Rheology}
\label{sec:linear-rheology}

For small shear rate $\dot\gamma\to0$ and packing fractions $\varphi <
\varphi_c$ below the glass transition, we recover the linear response of a
Newtonian fluid. We consider two values for the Hertzian stiffness $\Gamma =
1000$ and $\Gamma = 1500$ as well as several values for the coefficient of
restitution $\epsilon$. For the packing fraction we choose values high enough
for the validity of our approach but well below the hard-sphere random close
packing density $\varphi_J \approx 0.64$ \cite{torquato+stillinger10} such
that we avoid lasting contacts between particles.

The \textsc{ITT} formalism yields the following expression for the
viscosity 
\begin{equation}
  \label{eq:2}
  \eta = \eta_0 + \frac{3}{10\pi^2}\times\frac{(1 + \epsilon)^2}{4}
  \times\frac{T}{dD_S}\varphi^2\chi^2\tilde\eta[S](cd/D_S)
\end{equation}
where $\chi$ is the pair correlation function at contact, $c$ denotes
the (long wavelength) speed of sound, and\footnote{Here $j_0(x)$ is
  the zeroth order spherical Bessel function.}
\begin{equation}
  \label{eq:3}
  \tilde\eta[S](K) = \int_0^{\infty}dk\frac{K^2 + 2k^2}{K^2 + k^2}
  \times\frac{[j''_0(k)]^2}{S_k^2}.
\end{equation}
We expect that generally the viscosity will be lower for more
dissipative particles (smaller $\epsilon$) but note that $D_S$
strongly depends on $\epsilon$ especially for nearly elastic particles
\cite{garzo+montanero02}. The influence of the particles stiffness
$\Gamma$ will enter through the static structure factor $S_k$.

\begin{figure}
  \centering
  \includegraphics{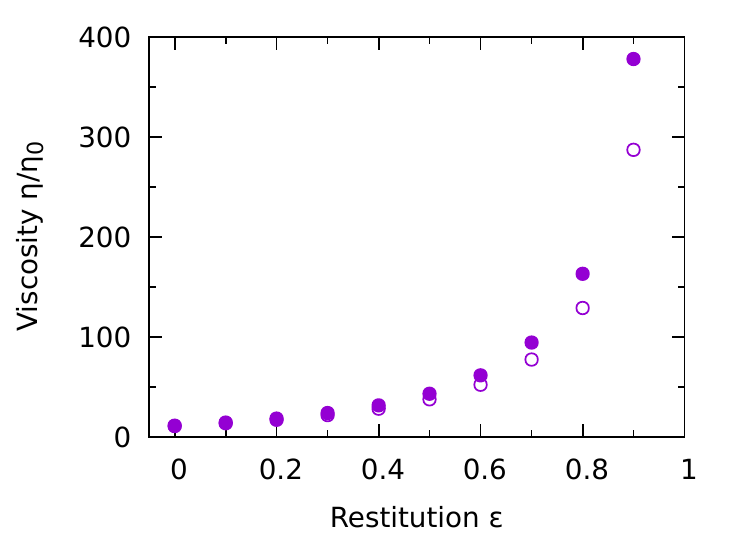}
  \caption{Viscosity $\eta$ relative to its low density limit $\eta_0$
    as a function of the coefficient of restitution $\epsilon$ for
    packing fraction $\varphi = 0.51$ and stiffness $\Gamma=1000$
    (open), and $1500$ (filled).}
  \label{fig:eta1}
\end{figure}

For moderate densities (cf.\ Fig.~\ref{fig:eta1}), we observe that the
viscosity increases with $\epsilon$ as expected. Note that we obtain
values that are far above the Boltzmann-Enskog value $\eta_0$ valid
for vanishing density. Also the stiffer particles have a higher
viscosity. This increase is explained by the approach to the glass
transition $\varphi_c(\Gamma)$ which is lower for stiffer particles
\cite{berthier+moreno10,heinen+sperl16}.

\begin{figure}
  \centering
  \includegraphics{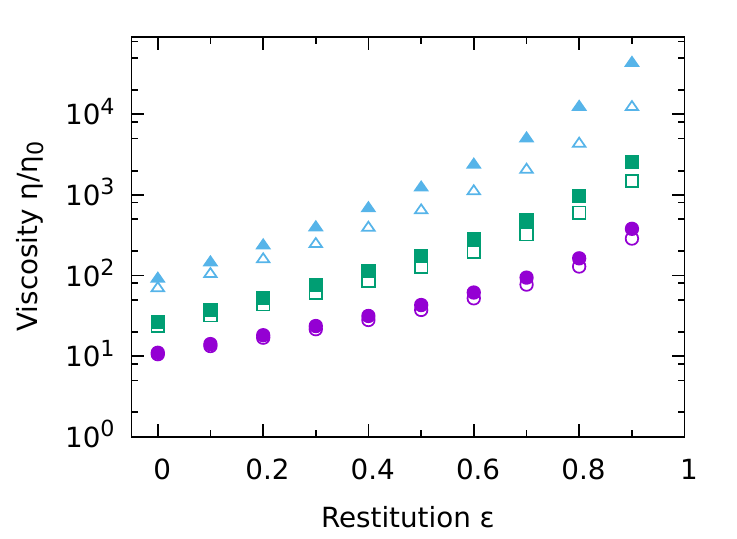}
  \caption{Viscosity $\eta$ relative to its low density limit $\eta_0$
    as a function of the coefficient of restitution $\epsilon$ for
    packing fractions $\varphi = 0.51$ (violet disks), $0.56$ (green
    squares), and $0.61$ (blue triangles) and stiffness $\Gamma=1000$
    (open), and $1500$ (filled).}
  \label{fig:eta2}
\end{figure}

In general, the viscosity increases over several orders of magnitude upon
increasing the density (cf.\ Fig.~\ref{fig:eta2}. Note that we span more than
two orders of magnitude in viscosity by varying the inelasticity $\epsilon$ at
the highest packing fraction considered.

\section{Nonlinear Rheology}
\label{sec:nonlinear-rheology}

For strong shear such that shear heating dominates the fluidization,
$\sigma\dot\gamma \gg nP_D$, we obtain Bagnold scaling where the
fluidized bed no longer behaves like a Newtonian fluid,
$\sigma = \eta\dot\gamma$, but $\sigma = B{\dot\gamma}^2$. This
peculiar behavior of granular fluids has first been observed and
described by Bagnold \cite{bagnold54}. Kinetic predictions for the
generalized (Bagnold) viscosity $B$ (which has dimensions of an
inverse length) are sparse.

The \textsc{ITT} formalism sketched above yields an expression for the
Bagnold coefficient \cite{kranz+frahsa16} in terms of the coherent
scattering function. Numerical evaluation of the resulting generalized
Green-Kubo-integrals yields values for the Bagnold coefficient $B$.

\begin{figure}
  \centering
  \includegraphics{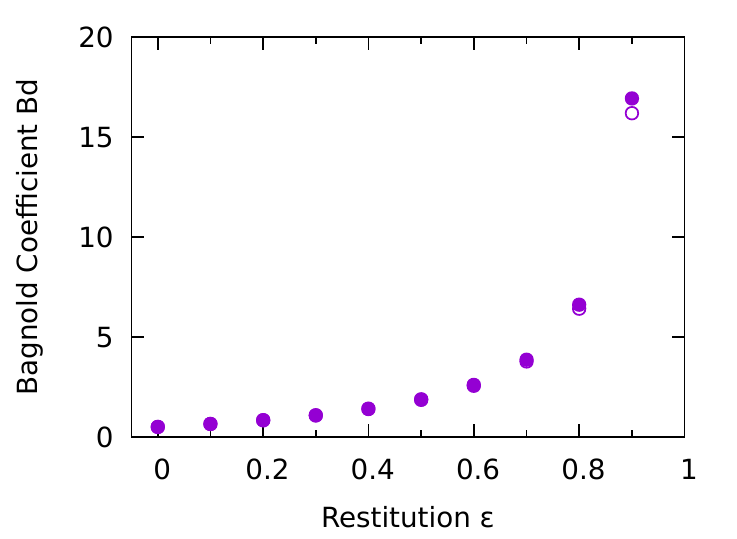}
  \caption{Bagnold coefficient $B\cdot d$ (where $d$ is the particle
    diameter) as a function of the coefficient of restitution
    $\epsilon$ for packing fraction $\varphi = 0.51$ and two values of
    the stiffness $\Gamma = 1000$ (open), and $1500$ (filled).}
  \label{fig:bag1}
\end{figure}

\begin{figure}
  \centering
  \includegraphics{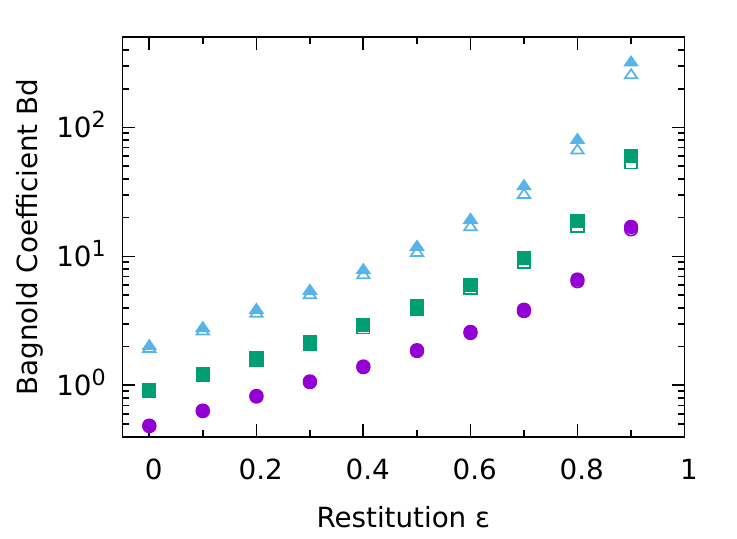}
  \caption{Bagnold coefficient $B\cdot d$ as a function of the
    coefficient of restitution $\epsilon$ for packing fractions
    $\varphi = 0.51$ (violet disks), $0.56$ (green squares), and
    $0.61$ (blue triangles) and two values of the stiffness
    $\Gamma = 1000$ (open), and $1500$ (filled).}
  \label{fig:bag2}
\end{figure}

We observe (cf.\ Fig.~\ref{fig:bag1}) a very weak dependence of $B$ on
the stiffness $\Gamma$ but, just as for the viscosity, a strong
increase with the coefficient of restitution $\epsilon$. As is
expected also the Bagnold viscosity increases with increasing
density (cf.\ Fig.~\ref{fig:bag2})

\section{Discussion}
\label{sec:discussion}

We presented first results from an granular \textsc{ITT} formalism for
strongly sheared dense granular fluids. It yields kinetic theory expressions
for the viscosity beyond the lowest order in density. These allowed us to
calculate viscosities well above the Boltzmann viscosity $\eta_0$ and to
determine the dependence on the inelasticity (quantified by the coefficient of
restitution $\epsilon$) and on the parameter $\Gamma$ controlling the
stiffness of the Hertzian potential. The general result is that the more
elastic and the stiffer the particles are, the higher is the viscosity of the
granular fluid.

We also presented predictions for the Bagnold coefficient $B$ which determines
the non-Newtonian rheology of granular hard spheres at high shear rates. Like
the Newtonian viscosity, it is a measure of the sluggishness of the fluid and
thus increases towards the glass transition packing fraction of the quiescent
fluid. We found that $B$ is rather insensitive to the stiffness $\Gamma$.

In the future we would expect our constitutive relations to be used in the
modeling of dense granular at practically relevant shear rates. Kumaran
\cite{kumaran14} recently presented work in this direction.

\section*{Acknowledgments}
We acknowledge continuing discussions with Annette Zippelius, Matthias
Fuchs, and Fabian Frahsa and thank Marco Heinen for the structure
factor code. W. T. K. acknowledges financial support by FOR1394.

%

\end{document}